# Light tunneling inhibition in array of couplers with combined longitudinal modulation of refractive index

Yaroslav V. Kartashov[1] and Victor A. Vysloukh[2]
[1]*ICFO-Institut de Ciencies Fotoniques, and Universitat Politecnica de Catalunya, Mediterranean Technology Park, 08860 Castelldefels (Barcelona), Spain*
[2]*Departamento de Fisica y Matematicas, Universidad de las Americas – Puebla, Santa Catarina Martir, 72820, Puebla, Mexico*

We consider light tunneling inhibition in periodic array of optical couplers due to the specially designed longitudinal and transverse modulation of the refractive index. We show that local out-of-phase longitudinal modulation of refractive index in channels of directional couplers in combination with the global refractive index modulation between adjacent couplers allow simultaneous suppression of both local and global energy tunneling inside each coupler and between adjacent couplers. This enables localization of light in single waveguide despite the remarkable difference of corresponding local and global energy tunneling rates.

Waveguide arrays offer exceptional opportunities for control of light propagation [1,2]. Additional degrees of freedom appear if the refractive index vary also in the direction of light propagation. Such structures are capable to support discrete diffraction-managed solitons [3-5] and allow flexible control of beam propagation direction [6,7]. Dynamic localization of light in photonic structures with longitudinal modulation of guiding parameters undoubtedly is among the most interesting optical phenomena. Such localization was predicted and observed in waveguide arrays [8-13] and optical couplers [14-17]. Different tools for the control of light tunneling were developed, such as periodic bending [8-10,14,15] or out-of-phase modulation of refractive index of adjacent guides [11-13,17]. All previous efforts were focused on photonic structures with a single characteristic energy exchange scale between neighboring guides. However the periodic array of optical couplers serves as an illuminating example of photonic structure where rapid local energy exchange between guides in each coupler coexists with slow global energy tunneling into adjacent couplers. The presence of two characteristic energy tunneling scales in this system makes the problem of light localiza-tion in a single channel especially challenging, since one may expect that simple longitudinal modulation usually adopted in waveguide arrays will not yield inhibition of tunneling. Note that potential analogy between optical and quantum problems [18] broadens the interest to this optical setting, which is similar to one-dimensional dimer model of quantum mechanics [19].

In this Letter we report on light tunneling inhibition in a periodic array of optical couplers due to specially designed longitudinal and transverse modulations of the refractive index. We show that by properly selecting the law of longitudinal modulation a relatively slow energy exchange between adjacent couplers in the array and rapid energy exchange between channels of individual couplers can be inhibited simultaneously that results in diffractionless propagation of single-channel excitations at certain resonant values of modulation frequency.

The propagation of laser radiation along the $\xi$-axis of periodic array of couplers is described by the Schrödinger equation for the normalized complex field amplitude $q$:

$$i\frac{\partial q}{\partial \xi} = -\frac{1}{2}\frac{\partial^2 q}{\partial \eta^2} - pR(\eta,\xi)q, \tag{1}$$

where $\eta$ and $\xi$ are the transverse and longitudinal coordinates normalized to the characteristic transverse scale and diffraction length, respectively, while the parameter $p$ describes refractive index modulation depth. The global refractive index distribution is described by the function $R = \sum_{k=-\infty}^{+\infty} R_k$, where refractive index profile of $k$-th coupler is given by

$$\begin{aligned} R_k = [1 + \mu_\mathrm{s}\sin(\Omega_\mathrm{s}\xi) \pm \mu_\mathrm{l}\sin(\Omega_\mathrm{l}\xi)]G(\eta + w_\mathrm{s}/2 + kw_\mathrm{l}) + \\ [1 - \mu_\mathrm{s}\sin(\Omega_\mathrm{s}\xi) \pm \mu_\mathrm{l}\sin(\Omega_\mathrm{l}\xi)]G(\eta - w_\mathrm{s}/2 + kw_\mathrm{l}), \end{aligned} \tag{2}$$

where $G(\eta) = \exp(-\eta^6 / w_\eta^6)$, $w_s$ is the separation between channels inside each coupler, $w_\eta$ is the channel width, $w_\mathrm{l}$ stands for the distance between centers of couplers, while signs $\pm$ correspond to even/odd values of $|k|$. This profile corresponds to the local out-of-phase harmonic modulation of the refractive index inside two channels of each coupler with a spatial frequency $\Omega_\mathrm{s}$ and modulation depth $\mu_\mathrm{s}$, and simultaneous global out-of-phase modulation of refractive index between neighboring couplers with different spatial frequency $\Omega_\mathrm{l}$ and modulation depth $\mu_\mathrm{l}$ [see Fig.1(a) for representative example of such an array of couplers]. Further we set $w_\eta = 0.3$, $w_\mathrm{s} = 1.6$, $w_\mathrm{l} = 4.0$, and $p = 7$. Notice that longitudinal refractive index modulation changes not only propagation constants of guided modes but it also modi-fies coupling between waveguides due to modification of guided mode profiles and overlap integrals that affect the rate of diffraction in the structure.

Figure 1(b) shows usual diffraction spreading for the case of excitation of single channel in central coupler in the absence of longitudinal modulation. We use linear guided mode of a single isolated guide as an initial condition for numerical integration of Eq. (1) (other input beam shapes yield qualitatively similar results). One can observe fast local oscillations inside individual couplers (spatial frequency of intensity beatings is given by $\Omega_{\mathrm{b}} = 2\pi / T_{\mathrm{b}}$, where for our parameters $T_{\mathrm{b}} = 21.6$) and slow global light spreading across the array due to light tunneling between adjacent couplers. Figure 1(c) illustrates an attempt of light tunneling inhibition using out-of-phase refractive index modulation only inside individual couplers ($\mu_{\mathrm{s}} \neq 0$, $\mu_{\mathrm{l}} = 0$). Notice that the modulation frequency $\Omega_{\mathrm{s}}$ and depth $\mu_{\mathrm{s}}$ selected correspond to the optimal light tunneling inhibition in an isolated coupler. While such modulation drastically suppresses energy exchange inside couplers (note that suppression is remarkable, but it can not be complete) it can not suppress light tunneling to neighboring couplers. Analogously, if only modulation between couplers is present ($\mu_{\mathrm{s}} = 0$, $\mu_{\mathrm{l}} \neq 0$), one can effectively suppress energy exchange between the couplers but not beatings inside input coupler [Fig. 1(d)]. This indicates that simple single-frequency longitudinal refractive index modulation does not allow to achieve tunneling inhibition in system with two characteristic energy tunneling scales and one has to resort to more complicated simultaneous local and global longitudinal modulation of refractive index, such as the one described by Eq. (2) with $\mu_{\mathrm{s}}, \mu_{\mathrm{l}} \neq 0$. The key issue is thus optimal selection of parameters of such a modulation.

As a criterion of optimization we used the distance-averaged energy flow $U_{1\mathrm{m}}$ trapped in the input channel of central coupler as well as distance-averaged energy flow $U_{2\mathrm{m}}$ in the entire central coupler:

$$\begin{aligned} U_{1\mathrm{m}} &= L^{-1}\int_0^L d\xi \int_{-w_{\mathrm{s}}}^{0} |q(\eta,\xi)|^2\, d\eta \Big/ \int_{-w_{\mathrm{s}}}^{0} |q(\eta,0)|^2\, d\eta, \\ U_{2\mathrm{m}} &= L^{-1}\int_0^L d\xi \int_{-w_{\mathrm{l}}/2}^{w_{\mathrm{l}}/2} |q(\eta,\xi)|^2\, d\eta \Big/ \int_{-w_{\mathrm{l}}/2}^{w_{\mathrm{l}}/2} |q(\eta,0)|^2\, d\eta, \end{aligned} \tag{3}$$

A characteristic feature of light tunneling inhibition is the resonant behavior of the distance-averaged energy flow trapped in the input channel of central coupler $U_{1\mathrm{m}}(\Omega_{\mathrm{s}})$ as well as distance-averaged energy in the entire central coupler $U_{2\mathrm{m}}(\Omega_{\mathrm{l}})$, as shown in Fig. 2. Panel 2(a) depicts $U_{1\mathrm{m}}$ dependence on the frequency $\Omega_{\mathrm{s}}$ for nonzero $\mu_{\mathrm{s}}$ in the absence of global modu-lation, while panel 2(b) illustrates $U_{2\mathrm{m}}$ dependence on $\Omega_{\mathrm{l}}$ for nonzero $\mu_{\mathrm{l}}$ in the absence of local modulation. Notice that the resonant peaks in the $U_{1\mathrm{m}}(\Omega_{\mathrm{s}})$ dependence are remarkably sharper. Interestingly, the frequencies of primary resonances (i.e., resonances that occur for largest $\Omega_{\mathrm{s}}$ or $\Omega_{\mathrm{l}}$ values) do not differ considerably for $U_{1\mathrm{m}}(\Omega_{\mathrm{s}})$ and $U_{2\mathrm{m}}(\Omega_{\mathrm{l}})$ dependencies from Fig. 2. Figure 2(c) exemplify almost linear dependencies of the principal resonance frequency $\Omega_{\mathrm{r}1}$ and the frequency of second resonance $\Omega_{\mathrm{r}2}$ on the depth of global modulation $\mu_{\mathrm{l}}$ at $\mu_{\mathrm{s}} = 0$. It should be stressed that the resonances in Figs. 2(a) and 2(b) correspond to suppression of coupling either inside couplers or between couplers, but overall localization that can be characterized by the combined localization criterion $U_{\mathrm{tm}} = U_{1\mathrm{m}} U_{2\mathrm{m}}$ remains low (i.e. $U_{\mathrm{tm}}$ is considerably smaller than 1) in both cases because only one type of modulation is present.

Figure 3 demonstrates combined localization criterion as a function of modulation frequencies $\Omega_{\mathrm{s}}$ and $\Omega_{\mathrm{l}}$ for simultaneous local and global longitudinal modulation of refractive index for the case when $\mu_{\mathrm{s}} \neq \mu_{\mathrm{l}}$ [Fig. 3(a)] and when $\mu_{\mathrm{s}} = \mu_{\mathrm{l}}$ [Fig. 3(b)]. The combined localization criterion $U_{\mathrm{tm}}$ attains a maximal value when energy tunneling between channels of individual coupler and between neighboring couplers is suppressed simultaneously. Importantly, the strongest overall localization with $U_{\mathrm{tm}} \approx 1$ (or the principal resonance for combined modulation) always appears in the vicinity of $\Omega_{\mathrm{s}}$ value that corresponds to light tunneling inhibition inside individual couplers and even small detuning of $\Omega_{\mathrm{s}}$ from this resonant value leads to remarkable diminishing of localization. The $U_{\mathrm{tm}}(\Omega_{\mathrm{l}})$ dependence is characterized by the narrow bands of strong delocalization around the frequencies $\Omega_{\mathrm{l}} = \Omega_{\mathrm{s}} / n$, where $n \geq 1$ is an integer, while maxima of $U_{\mathrm{tm}}$ corresponding to overall inhibition of tunneling appear exactly in between these delocalization bands. For a fixed $\Omega_{\mathrm{s}}$ the comparable inhibition of tunneling can be achieved for several $\Omega_{\mathrm{l}}$ values. Figure 3(c) illustrates the dependence of combined localization criterion on the local modulation depth $\mu_{\mathrm{s}}$ when $\mu_{\mathrm{l}}$, $\Omega_{\mathrm{s}}$, and $\Omega_{\mathrm{l}}$ are fixed and correspond to strongest global resonance. Typical features of $U_{\mathrm{tm}}(\mu_{\mathrm{s}})$ profile are sharp localization peak and well defined localization minimum which appear due to almost linear dependence of resonant frequencies on the modulation depth $\mu_{\mathrm{s}}$ [see also Figs. 2(a)-2(c)]. The examples of complete light tunneling inhibition are shown in Figs. 1(e) and 1(f) for the case of combined longitudinal modulation with $\mu_{\mathrm{s}}, \mu_{\mathrm{l}} \neq 0$. One can see that by properly selecting corresponding modulation frequencies $\Omega_{\mathrm{s}}$ and $\Omega_{\mathrm{l}}$ one can simultaneously suppress energy exchange between channels of coupler and between neighboring couplers, so that upon propagation the light remains in the excited channel.

Summarizing, we showed that specially designed out-of-phase longitudinal modulation of refractive index in the channels of directional coupler in combination with modulation of the refractive index between adjacent couplers allow precise control of tunneling despite remarkable difference of corresponding energy exchange scales. This may be useful for creation of multichannel optical couplers where output patterns can be altered dramatically by only slight modifications in longitudinal refractive index modulation frequency and where even weak nonlinearity may substantially affect switching dynamics.

## References with titles

## References without titles

## Figure captions

Figure 1. (a) An example of $\xi$-modulated array of couplers. (b) Discrete diffraction in unmodulated array. Propagation dynamics in modulated array at (c) $\mu_s = 0.2$, $\mu_l = 0$, $\Omega_s = 2.77\Omega_b$, (d) at $\mu_s = 0$, $\mu_l = 0.2$, $\Omega_l = 2.84\Omega_b$, (e) at $\mu_s = 0.2$, $\mu_l = 0.1$, $\Omega_s = 2.77\Omega_b$, $\Omega_l = 1.23\Omega_b$, and (f) at $\mu_s = \mu_l = 0.15$, $\Omega_s = 2.09\Omega_b$, $\Omega_l = 0.91\Omega_b$. The propagation distance in (b) is $L = 15T_b$, while in (c)-(f) it is $L = 30T_b$. White ticks at the bottom of each image indicate centers of couplers.

Figure 2. (a) Distance-averaged energy flow in the input channel versus $\Omega_s$ at $\mu_s = 0.2$, $\mu_l = 0$. (b) Distance-averaged energy flow in the input coupler versus $\Omega_l$ at $\mu_s = 0$, $\mu_l = 0.2$. (c) Frequencies of first and second resonances versus $\mu_l$ for $\mu_s = 0$.

Figure 3. The product of distance-averaged energy flows $U_{tm} = U_{1m}U_{2m}$ versus modulation frequencies $\Omega_s$ and $\Omega_l$ at (a) $\mu_s = 0.20$, $\mu_l = 0.10$, and (b) $\mu_s = 0.15$, $\mu_l = 0.15$. Red regions corresponds to strongest inhibition of tunneling when $U_m \approx 1$, while in blue regions one has strongest delocalization when $U_m \to 0$. (c) $U_{tm}$ versus $\mu_s$ at $\mu_l = 0.1$, $\Omega_s = 2.77\Omega_b$, $\Omega_l = 1.23\Omega_b$.

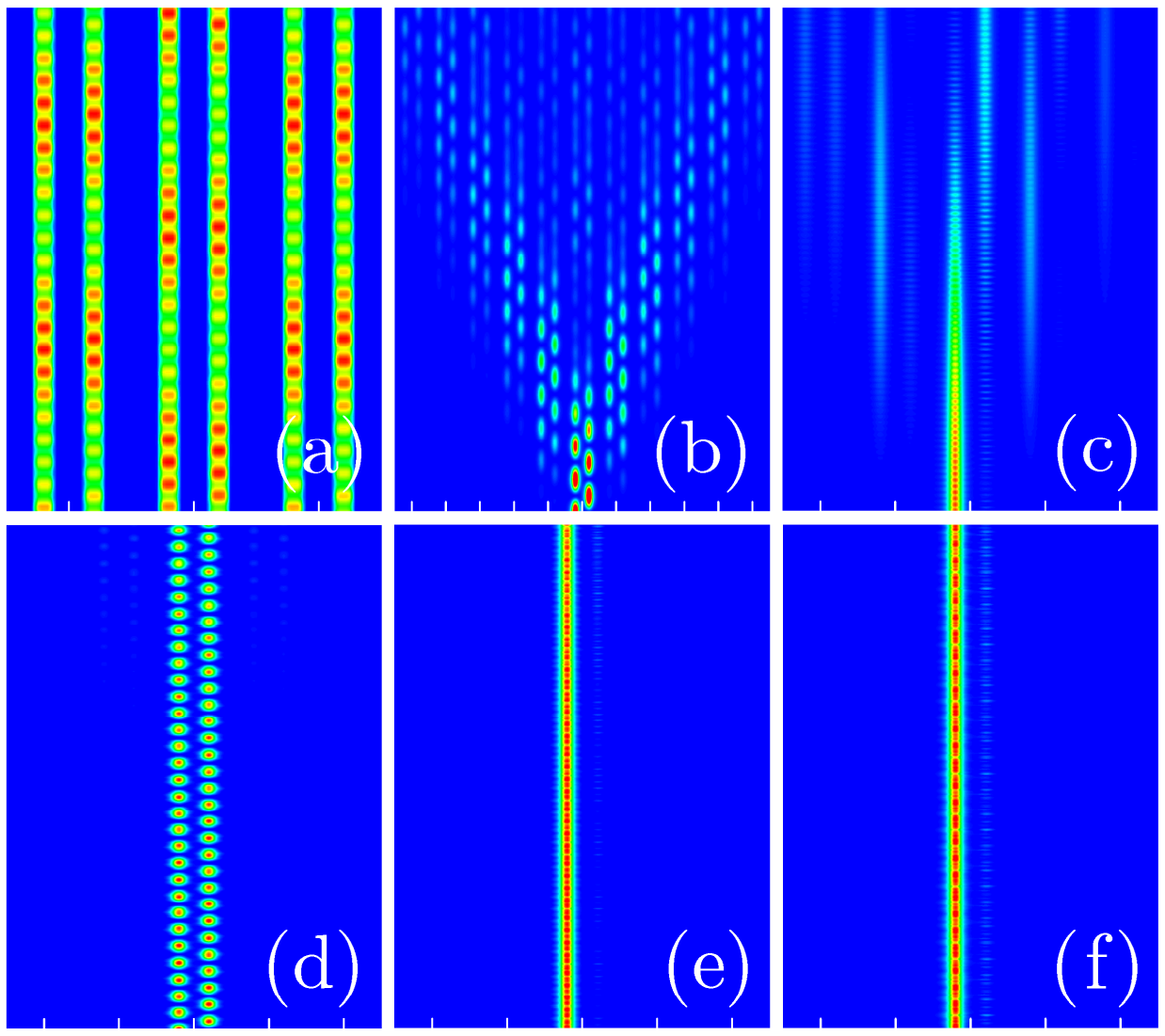

Figure 1. (a) An example of $\xi$ -modulated array of couplers. (b) Discrete diffraction in unmodulated array. Propagation dynamics in modulated array at (c) $\mu_{\rm s} = 0.2$ , $\mu_{\rm l} = 0$ , $\Omega_{\rm s} = 2.77\Omega_{\rm b}$ , (d) at $\mu_{\rm s} = 0$ , $\mu_{\rm l} = 0.2$ , $\Omega_{\rm l} = 2.84\Omega_{\rm b}$ , (e) at $\mu_{\rm s} = 0.2$ , $\mu_{\rm l} = 0.1$ , $\Omega_{\rm s} = 2.77\Omega_{\rm b}$ , $\Omega_{\rm l} = 1.23\Omega_{\rm b}$ , and (f) at $\mu_{\rm s} = \mu_{\rm l} = 0.15$ , $\Omega_{\rm s} = 2.09\Omega_{\rm b}$ , $\Omega_{\rm l} = 0.91\Omega_{\rm b}$ . The propagation distance in (b) is $L = 15T_{\rm b}$ , while in (c)-(f) it is $L = 30T_{\rm b}$ . White ticks at the bottom of each image indicate centers of couplers.

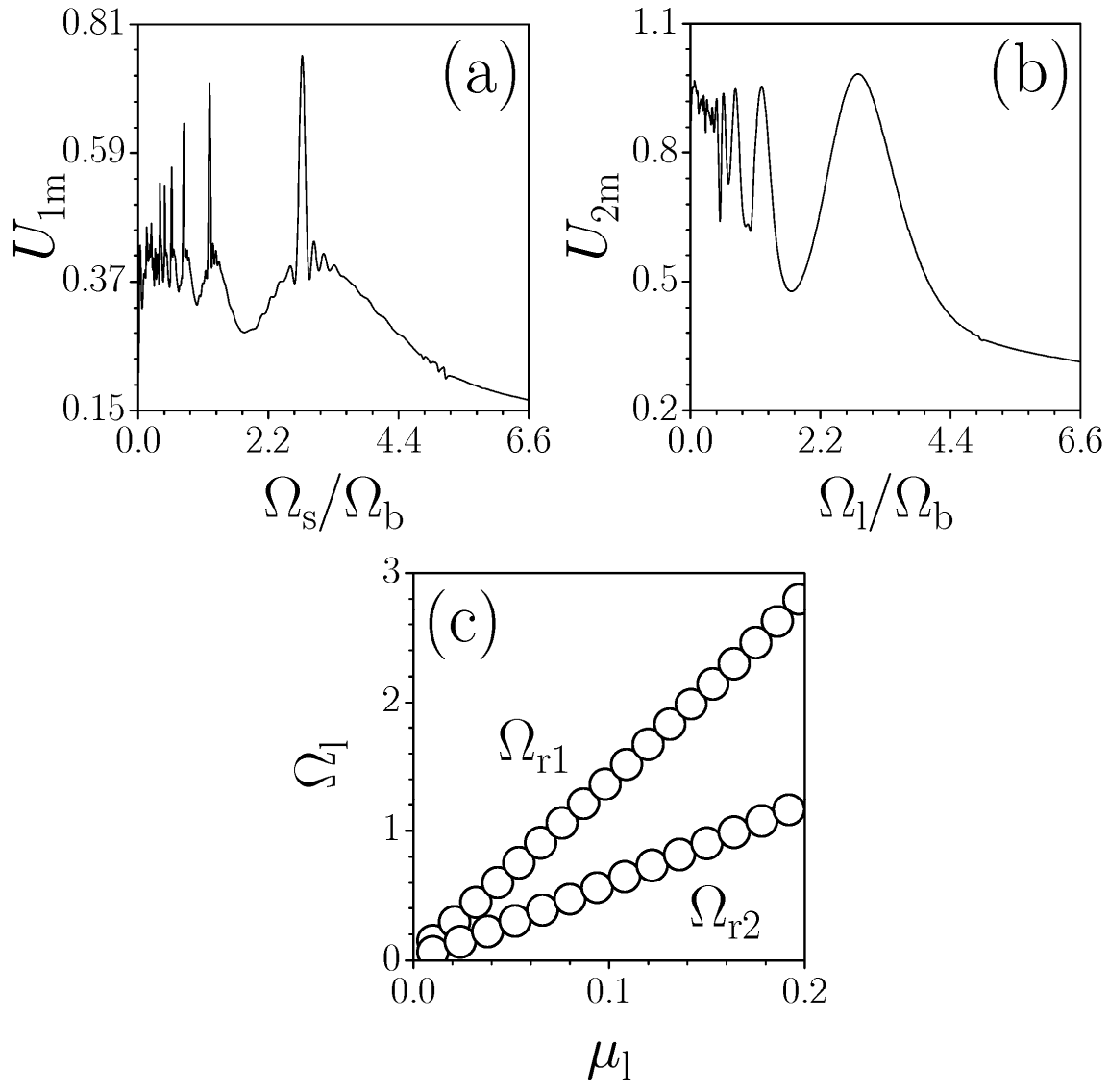

Figure 2. (a) Distance-averaged energy flow in the input channel versus $\Omega_s$ at $\mu_s = 0.2$, $\mu_l = 0$. (b) Distance-averaged energy flow in the input coupler versus $\Omega_l$ at $\mu_s = 0$, $\mu_l = 0.2$. (c) Frequencies of first and second resonances versus $\mu_l$ for $\mu_s = 0$.

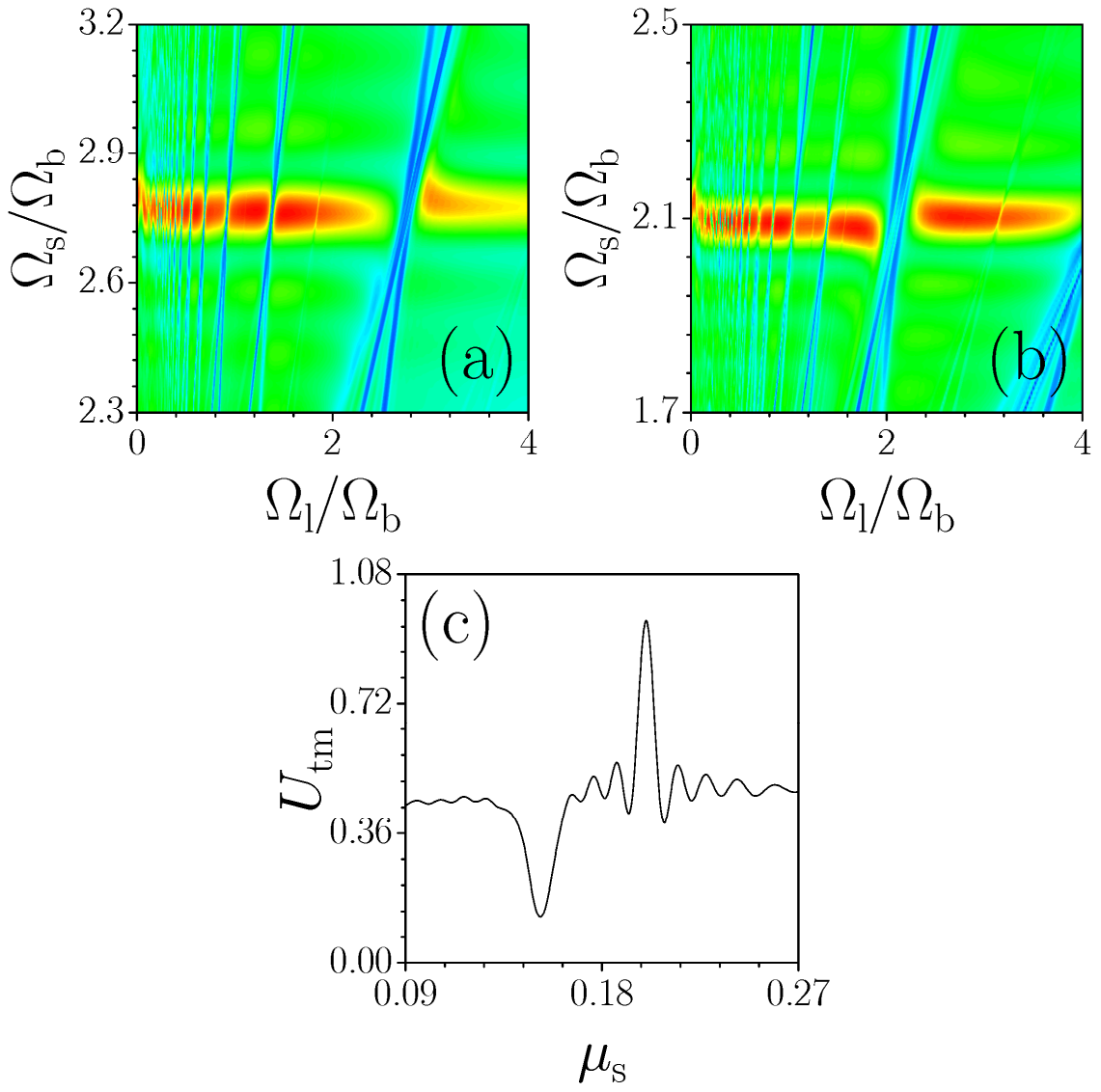

Figure 3. The product of distance-averaged energy flows $U_{\mathrm{tm}} = U_{1\mathrm{m}}U_{2\mathrm{m}}$ versus modulation frequencies $\Omega_{\mathrm{s}}$ and $\Omega_{\mathrm{l}}$ at (a) $\mu_{\mathrm{s}} = 0.20$, $\mu_{\mathrm{l}} = 0.10$, and (b) $\mu_{\mathrm{s}} = 0.15$, $\mu_{\mathrm{l}} = 0.15$. Red regions corresponds to strongest inhibition of tunneling when $U_{\mathrm{m}} \approx 1$, while in blue regions one has strongest delocalization when $U_{\mathrm{m}} \to 0$. (c) $U_{\mathrm{tm}}$ versus $\mu_{\mathrm{s}}$ at $\mu_{\mathrm{l}} = 0.1$, $\Omega_{\mathrm{s}} = 2.77\Omega_{\mathrm{b}}$, $\Omega_{\mathrm{l}} = 1.23\Omega_{\mathrm{b}}$.